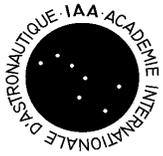
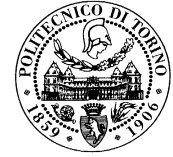

**INTERNATIONAL ACADEMY OF ASTRONAUTICS**
8th IAA SYMPOSIUM ON THE FUTURE OF SPACE
EXPLORATION: TOWARDS THE STARS

Torino, Italy,   July 3-5, 2013

# UNITED NATIONS HUMAN SPACE TECHNOLOGY INITIATIVE (HSTI)

**Mika OCHIAI, Aimin NIU, Heike STEFFENS, Werner BALOGH,
Hans HAUBOLD, Mazlan OTHMAN, Takao DOI**
United Nations Office for Outer Space Affairs, Wagramer Strasse 5, 1400 Vienna, Austria
hsti@unoosa.org

**ABSTRACT**

The Human Space Technology Initiative (HSTI) was launched in 2010 within the framework of the United Nations Programme on Space Applications, aimed at involving more countries in activities related to human spaceflight and space exploration and at increasing the benefits from the outcome of such activities through international cooperation, to make space exploration a truly international effort.

The role of the Initiative in these efforts consists of providing a platform to exchange information, foster collaboration between partners from space-faring and non-space-faring countries, and encourage emerging and developing countries to take part in space research and benefit from space applications.

The Initiative organizes expert meetings and workshops annually to raise awareness of the current status of space exploration activities, as well as of the benefits of utilizing human space technology and its applications. In close cooperation with the International Space Station (ISS) partners, information on the ISS has been provided. In 2013, the United Nations/China Workshop on Human Space Technology in Beijing, China, from 16 to 20 September 2013 took place as part of HSTI, hosted by the China Manned Space Agency and co-organized by the International Academy of Astronautics (IAA).

HSTI has been carrying out its primary science activity called the Zero-Gravity Instrument Project (ZGIP) and a new fellowship programme the "Drop Tower Experiment Series" (DropTES). Those are part of the efforts of promoting capacity-building activities in microgravity science and education, particularly in developing countries. Under the Zero-Gravity Instrument Project, the Office for Outer Space Affairs provides clinostats (microgravity simulation instruments) to selected schools and universities to perform experiments at their own Earth laboratories.

## 1. INTRODUCTION

Space exploration fires people's imaginations. Since the first human spaceflight in 1961, over 500 explorers from different nations have ventured into space, motivated by curiosity, the drive for knowledge, and the belief that space exploration could benefit people on Earth. The involvement of a growing number of countries means that space exploration and the use of outer space are now truly global undertakings. Given the great importance of international cooperation in the peaceful exploration and use of outer space, the United Nations Committee on the Peaceful Uses of Outer Space (COPUOS) fills the need for an intergovernmental platform at the global level in this regard.

In the Declaration on the Fiftieth Anniversary of Human Space Flight and the Fiftieth Anniversary of the Committee on the Peaceful Uses of Outer Space, adopted by the United Nations General Assembly in 2011 [1], States emphasized the significant progress in the development of space science and technology and their applications that had enabled humans to explore the universe, and the extraordinary achievements made over the previous 50 years in space exploration efforts, including deepening the understanding of the planetary system and the Sun and the Earth itself, in the use of space science and technology for the benefit of all humankind and in the development of the international legal regime governing space activities.

The United Nations Office for Outer Space Affairs (OOSA), as the secretariat of COPUOS, services international discourses and dialogues against a backdrop of law and diplomacy, conducts capacity-building activities that harness space tools for development and promotes awareness through the celebration of space achievements and milestones. It works to meet the needs of developing countries, to advance the space-faring technologies of nations and to encourage the emergence of new actors in space. The Human Space Technology Initiative, an activity conducted under the United Nations Programme on Space Applications, is intended to help more countries benefit from space exploration and human space technology, and to open opportunities for further engagement in space exploration.

## 2. UNITED NATIONS AND HUMAN SPACEFLIGHT

In the 2011 Declaration on the Fiftieth Anniversary of Human Space Flight and the Fiftieth Anniversary of the Committee on the Peaceful Uses of Outer Space, States recalled the launch into outer space of the first human-made Earth satellite, Sputnik I, on 4 October 1957, thus opening the way for space exploration; also recalled that, on 12 April 1961, Yuri Gagarin became the first human to orbit the Earth, opening a new chapter of human endeavour in outer space; and further recalled the amazing history of human presence in outer space and the remarkable achievements since the first human spaceflight, in particular, Valentina Tereshkova becoming the first woman to orbit the Earth on 16 June 1963, Neil Armstrong becoming the first human to set foot upon the surface of the Moon on 20 July 1969, and the docking of the Apollo and Soyuz spacecraft on 17 July 1975, being the first international human mission in space; and recalled that for the past decade humanity had maintained a multinational permanent human presence in outer space onboard the International Space Station (ISS).

In establishing COPUOS in 1959, the General Assembly requested this body to review the scope of international cooperation in peaceful uses of outer space and to devise programmes in that field, which could be undertaken under United Nations auspices, to encourage continued research and the dissemination of information on outer space matters, as well as to study legal problems arising from the exploration of outer space [2]. OOSA and its predecessors within the Secretariat of the United Nations have been assisting the Committee and its Subcommittees to achieve their goals and implement their decisions since the late 1950s.

There are several examples of the achievements of the Committee over the past 50 years for which the establishment of the legal regime governing the exploration and use of outer space is of fundamental importance. The first United Nations treaty on outer space, the Treaty on Principles Governing the Activities of States in the Exploration and Use of Outer Space, including the Moon and Other Celestial Bodies (Outer Space Treaty) of 1967, established, inter alia, that the "exploration and use of outer space, including the Moon and other celestial bodies, shall be carried out for the benefit and in the interests of all countries, irrespective of their degree of economic or scientific development, and shall be the province of all mankind" [3].

The Third United Nations Conference on the Peaceful Uses of Outer Space (UNISPACE III), held in 1999, recognized that large human space exploration missions exceeded the capacity of a single country and that cooperation should be privileged in that area. The ISS was cited as an example of that new paradigm which had been made possible by the end of the cold war [4]. UNISPACE III recommended the development of future space science programmes, in particular, through international cooperation and the encouragement of access to the ISS by countries that had never participated in that endeavor [5]. It also advocated the worldwide dissemination of information about research activities onboard the ISS. One year later, in 2000, a crew boarded the ISS for the first time.

The Scientific and Technical Subcommittee of COPUOS, at its 37th Session, in February 2000, began its consideration of a new agenda item for discussion entitled "International cooperation in human space flight", in accordance with General Assembly resolution 54/67. The Subcommittee had before it a working paper by the United States of America that presented an overview of the ISS. Delegations reviewed past, current and upcoming national and international programmes of cooperation in human spaceflight. Examples were given of international activities involving cooperation in national human spaceflight programmes such as Apollo, Soyuz, Salyut, Skylab, the Space Shuttle, Spacelab and the Mir Space Station. In addition, the nature and role of the ISS as well as the activities and contributions of various States leading up to its development, assembly and utilization were discussed.

In June 2007, during the fiftieth session of COPUOS, the High-level Panel on Space Exploration offered insights into ongoing national and global space exploration initiatives and the possible future role the United Nations system could play in providing a forum for space-faring as well as space-using, countries to consider issues related to space exploration.

In 2011, the General Assembly declared 12 April as the International Day of Human Space Flight, to celebrate each year at the international level the beginning of the space era for mankind, reaffirming the important contribution of space science and technology in achieving sustainable development goals and increasing the well-being of States and peoples, as well as ensuring the realization of their aspiration to maintain outer space for peaceful purposes.

In 2011, COPUOS marked the fiftieth anniversary of humankind's first flight into outer space. This commemoration resulted in the adoption of the Declaration on the Fiftieth Anniversary of Human Space Flight and the Fiftieth Anniversary of the Committee on the Peaceful Uses of Outer Space, in which States Members of the United Nations stressed the need to look more closely into how advanced space research and exploration systems and technologies could further contribute to meeting challenges, including that of global climate change, and to food security and global health, and endeavour to examine how the outcomes and spin-offs of scientific research in human space flight could increase the benefits, in particular for developing countries.

3. **UNITED NATIONS PROGRAMME ON SPACE APPLICATIONS**
The United Nations Programme on Space Applications, since its creation in 1971, has made substantial progress in fostering knowledge of and experience related to space applications around the world. The activities of the Programme are carried out by OOSA, with the annual endorsement of COPUOS. The mission of the Programme is to enhance the understanding and subsequent use of space technology for peaceful purposes, in general, and for national development in particular, in response to expressed needs in different geographic regions of the world.

In order to promote space science and technology and their applications, the Programme organized 296 workshops and seminars in 71 countries. About 20,000 experts from around the world participated in them from 1971 to 2013,

The overall strategy of the Programme is to focus on selected areas that are critical for developing countries, defining and working towards objectives achievable in two to five years and built on the results of previous activities. Those priority areas of the Programme are: (a) environmental monitoring; (b) natural resource management; (c) satellite communications for tele-education and telemedicine applications; (d) disaster risk reduction; (e) the use of global navigation satellite systems; (f) the Basic Space Science Initiative, including the International Space Weather Initiative; (g) space law; (h) climate change; (i) the Basic Space Technology Initiative; and (j) the Human Space Technology Initiative. Additional Programme areas include spin-offs of space technology and promotion of the participation of youth in space activities and of private industry in the activities of the Programme.

In response to the interest expressed by Member States, the Programme on Space Applications considered activities related to human spaceflight and exploration, which led to the launch of the Human Space Technology Initiative in 2010 with its long range plans and goals. HSTI and two other Initiatives, the Basic Space Science Initiative (BSSI) and the Basic Space Technology Initiative (BSTI), represent new cornerstones of the Programme on Space Applications [6].

4. **HUMAN SPACE TECHNOLOGY INITIATIVE**
HSTI aims at promoting space exploration as a common and unifying goal for humankind. Furthermore, increasing the use of space applications and spin-off technologies in emerging and developing countries as well as supporting the long-term sustainability effort of activities in space are also part of the mission of HSTI.

The role of HSTI in these efforts consists of providing a platform to exchange information, foster collaboration between partners from space-faring, and non-space-faring countries, and encourage emerging and developing countries to take part in space research and benefit from human space technology and its applications [7]. The activities of HSTI are based upon the following three pillars:

(a) **International Cooperation**: To promote international cooperation in human spaceflight and activities related to space exploration;
(b) **Outreach**: To promote increased awareness among Member States on the benefits of utilizing human space technology and its applications;
(c) **Capacity-Building**: To build capacity in microgravity science education and research.

The work of HSTI is based on the mandate of the United Nations Programme on Space Applications (General Assembly resolution 37/90, 10 December 1982). Each activity is in accordance with at least one element of the mandate.

8th IAA Symposium on the future of space exploration: towards the stars

**4.1 International Cooperation**

HSTI bridges and connects different partners from the international space community and other United Nations entities as well as Member States.

The International Space Station (ISS), being the largest permanently crewed research facility in space, has the inherent potential to benefit humankind in many ways. In cooperation with the ISS partners, HSTI provides information about the ISS, its management structure and its research facilities.

HSTI also informs Member States about opportunities to cooperate with space agencies and provides educational materials on space science and technology.

**4.1.1 Expanding the Benefits of the ISS for Humanity**

As the first activity of HSTI, OOSA organized a one-day activity called **"Outreach Seminar on the Activities of the International Space Station (ISS**)" in Vienna on 8 February 2011, during the 48th session of the Scientific and Technical Subcommittee of COPUOS [8]. With the purpose of disseminating information on the activities carried out onboard the ISS to a broader community as well as facilitating discussions on HSTI, the seminar was attended by a total of 17 countries including developing nations. The participants received first-hand information about the status of educational and research activities and about how to participate in research projects onboard the ISS. The seminar concluded that HSTI could be a meaningful mechanism for raising awareness of the potential of ISS research and educational activities among countries, regions, and potential users that had, up to that point, not been involved in such activities, thereby contributing to capacity-building in microgravity science and technology education.

Since the first Outreach Seminar, HSTI has focused on the potential benefits the ISS could offer to people on Earth, working together with the partner agencies of the ISS programme, namely the Canadian Space Agency (CSA), the European Space Agency (ESA), the Japan Aerospace Exploration Agency (JAXA), the National Aeronautics and Space Administration (NASA) of the United States of America, and the Russian Federal Space Agency (Roscosmos).

To facilitate cooperation in extending benefits of ISS research and education to the world, the **United Nations Expert Meeting on International Space Station Benefits for Humanity** was held from 11 to 12 June 2012 in Vienna, Austria, during the 49th session of COPUOS [9]. The meeting brought together experts from ISS partners and United Nations specialized agencies to discuss and identify potential collaborations, with the aim of extending the ISS benefits to all people in the identified areas of Earth observation and disaster response, health and education. Corresponding to these areas, the World Meteorological Organization (WMO), the United Nations Environment Programme (UNEP), the World Health Organization (WHO), and the United Nations Educational, Scientific and Cultural Organization (UNESCO) took part in the discussions.

The meeting served as groundwork for identifying potential ways of extending the benefits and follow-up activities in the areas of education and health are underway. As part of this follow-up, HSTI will organize an Expert Meeting with participants from the ISS partner agencies and WHO, especially tailored to the topic of health. The aim of the Expert Meeting, being scheduled for 19 and 20 February 2014 pursuant to General Assembly resolution 68/7, is to compile existing or new information related to the six leadership priorities of WHO, defined by the 66th World Health Assembly in its 12th General Programme of Work for the six-year period 2014–2019. The Expert Meeting will focus on facilitating a dialogue between ISS partner agencies and WHO to identify certain fields of collaboration where the needs and requirements of the health sector meet the offers deriving from space applications and technologies.

**4.1.2 Dialogue with Space Facility Owners – China's Space Station**

Over the past decade, with the growth of economic and technological development, an increasing number of emerging countries have shown interests and undertaken activities in human space exploration. China sent its first citizen into space in 2003 using its own spaceship. The construction of the Chinese space station is scheduled to begin in 2018 and become operational around 2020 [10].

During the 55th session of COPUOS in June 2012, the Government of China offered the utilization of the facilities on its planned manned space station to the world. Potential areas of cooperation through HSTI have been offered by the China Manned Space Agency (CMSA), and they are: technical cooperation on its construction; space experiments and applications; an international astronaut programme; and promotion of human space technology. HSTI has been working with CMSA to review possible collaboration in the utilization of China's space station to promote space exploration-related activities in the world.

### 4.1.3 Enhancing Synergies in Space Cooperation Efforts

HSTI has been engaging in discussions with various sectors to maximize existing efforts and complement them in fostering cooperative activities. In 2012, cooperation with the International Academy of Astronautics (IAA) was initiated in view of the common objective of fostering global cooperation in human spaceflight. Collaborative activities include supporting the United Nations workshop on human space technology in 2013, complementing HSTI projects and activities for capacity-building purposes in microgravity science, and providing opportunities for space experts from developing countries to take part in IAA's symposiums and conferences, including the IAA Thematic Summit on Planetary Robotics and Human Spaceflight and Head of Space Agencies which took place from 9 to 10 January 2014.

### 4.2 Outreach

HSTI organizes expert meetings and workshops annually to raise awareness of the current status of space exploration activities as well as of the benefits of utilizing human space technology and its applications. Experts from around the world meet together and exchange information and discuss possible future space projects with regard to human space exploration and its related activities. HSTI also provides publications and distributes informative materials on these subjects.

### 4.2.1 United Nations/Malaysia Expert Meeting on Human Space Technology

The **United Nations/Malaysia Expert Meeting on Human Space Technology** was held in Putrajaya, Malaysia, from 14 to 18 November 2011 [11]. This was the first United Nations event of its kind to discuss how the world can benefit from human space technology and to further develop international collaborative activities in human space exploration. Hosted by the Institute of Space Science of the National University of Malaysia and co-organized by OOSA and the five ISS partner agencies, 125 professionals from 23 countries participated in the meeting.

Throughout the five-day meeting, various presentations on the following topics were delivered: ISS programmes; microgravity science; education, outreach and capacity-building; and national, regional, and international space programmes. These were followed by discussions in three working group sessions: Microgravity Research; Education, Outreach and Capacity-Building; and HSTI. Participants provided remarks and observations on these themes with the final objective being to develop shared recommendations on the Initiative.

At the end of the meeting, ten recommendations were endorsed by the participants. The following five directly refer to the work of HSTI:

    (a) The Human Space Technology Initiative should take action to create awareness among stakeholders, including decision makers in the public and private sectors, researchers and students, of the social and economic potential of space science and technologies and to initiate outreach activities;

    (b) The Initiative should identify and inform Member States about space-related research opportunities and organize meetings in which invited experts can provide information to interested parties;

    (c) The Initiative should establish capacity-building programmes, including through the provision of educational materials, instrument distribution and/or access, national or regional expert centres, training of trainers, exchange programmes and competition and motivation programmes;

    (d) The Initiative should serve as a catalyst for international collaboration by promoting the formation of common interest groups, conducting regular surveys of countries concerning their space competence profiles, developing a set of guidelines for collaboration, promoting multi-national institutional agreements and creating regional expert centres;

    (e) The Initiative should promote the exchange of knowledge and the sharing of data by raising awareness, promoting user-friendly mechanisms for data access, and providing knowledge about self-supporting habitats for application, including for energy efficiency on Earth.

### 4.2.2 United Nations/China Workshop on Human Space Technology

As a further extension of the 2011 Expert Meeting, OOSA organized the United Nations/China Workshop on Human Space Technology in Beijing from 16 to 20 September 2013 [12]. The workshop was organized as part of HSTI within the framework of the Programme on Space Application 2013 activities, co-organized by the International Academy of Astronautics (IAA), and hosted by the China Manned Space Agency (CMSA) on behalf of the Government of China. The Workshop was attended by 150 professionals from governmental institutions, universities and other academic entities and non-governmental organizations from 31 countries.

With the aim of enabling participants to exchange information on human space exploration and to put forward constructive and innovative proposals on promoting international cooperation, the following recommendations were made:

(a) The Human Space Technology Initiative should play a role in informing Member States about the latest developments in human space exploration and in facilitating coordination among Member States to pursue common goals in a long-term perspective, identify opportunities for international cooperation and put forward proposals.

(b) The Initiative should promote education and outreach activities by providing educational materials as well as expert and astronaut forums to assist professionals and to inspire students, academia and the general public about human space exploration.

(c) Governments, institutions, industry and individuals are encouraged to participate in the global human space exploration endeavour. That would inspire young people by exposing them to new discoveries in science and technology and would enhance international cooperation in the pursuit of common goals of humanity.

(d) Governments and institutions are encouraged to create databases which include scientific, technical and legal information to promote the dissemination and exchange of information on human space exploration and its related activities.

(e) Governments and institutions are encouraged to establish educational mechanisms, develop appropriate curriculums and provide training for teachers in order to promote education in space science and technology.

## 4.3 Capacity-Building

HSTI is carrying out science activities aimed at contributing to promoting space education and research in microgravity around the world, particularly for enhancing capacity-building efforts in developing countries. Currently, HSTI focuses on promoting microgravity science education and research.

Everything that is living and non-living on Earth is exposed to gravity. All life forms evolved under the effects of this constant force. It is an essential factor and stimulus in most physical and physiological phenomena observed on our planet. For microgravity platforms such as drop towers, sounding rockets, spacecraft, or the International Space Station, however, the effects of this gravitational force is absent, and organisms experience microgravity conditions, which allows scientists to examine physical phenomena and the reaction of living cells, small organisms and even the human body in the absence of gravity. This unique environment and the scientific research making use of it will provide new insights into certain phenomena and processes. Understanding how organisms and matter react to gravity and the absence of it may also lead to new applications benefiting humankind. New medical cures and new materials may result from such knowledge. It may also open up the possibility for the human race to expand into space.

### 4.3.1 Zero-Gravity Instrument Project (ZGIP)

The "Zero-gravity Instrument Project (ZGIP)" provides unique opportunities for students and researchers to study the gravitational effects on samples, such as plant seeds and small organisms, in a simulated microgravity condition with hands-on learning in classroom or research activities conducted by each institution. Under this Project, as the first series of activities, a fixed number of microgravity simulation instruments - called Clinostats [13] - will be distributed free of charge to qualified schools, universities, research centers and institutes around the world [14].

ZGIP is aimed at motivating research institutions to invest in activities in space and microgravity research and fostering a global network of participating institutions in this field. ZGIP also expects to create a dataset of experimental results in gravity responses that could contribute to the design of future space experiments and to the advancement of microgravity research.

In order to select suitable institutions to receive the clinostats and increase the scientific value of the project, the HSTI Science Advisory Group (HSTI-SAG) was established and is comprised, on a voluntary basis, of prominent experts in microgravity life science. A "Teacher's Guide to Plant Experiments in Microgravity" was developed by OOSA with the support of the HSTI-SAG, and is electronically available on the project website [15]. The Guide provides straightforward instructions for teachers and students to perform experiments on plant growth using the clinostats in a school laboratory.

The Announcement of Opportunity of the first cycle was released on 1 February 2013. After the careful selection of 29 valid applications, a total of 19 schools and institutions from the following countries have been selected to take part in the project: Chile, China, Ecuador, Ghana, Iran, Iraq, Kenya, Malaysia, Nigeria, Pakistan, Thailand, and Viet Nam.

The announcement to call for applications for the second-cycle of ZGIP was released on 1 January 2014, and the application period is open until 30 April 2014, followed by a three-month evaluation phase of each submitted proposal and selection by HSTI-SAG and OOSA.

**4.3.2 Drop Tower Experiment Series (DropTES)**

The Drop Tower Experiment Series is a fellowship programme in which students can learn and study microgravity science by performing experiments in a drop tower. The Bremen Drop Tower is a ground-based laboratory with a drop tube height of 146 meters, which can enable short-term microgravity to be performed for various scientific fields such as fluid physics, combustion, thermodynamics, material science and biotechnology [16].

In collaboration with the Center of Applied Space Science and Microgravity (ZARM) and the German Aerospace Center (DLR), the fellowship programme offers a selected research team the opportunity to conduct its own microgravity experiments at the Drop Tower in Bremen, Germany. The series of experiments will consist of four drops or catapult launches during which approximately 5 to 10 seconds of microgravity, respectively, are produced.

The announcement to call for applications for the first-cycle of DropTES was released on 1 November 2013, and the application period is open until 28 February 2014 [17]. The first experiments by the selected team are expected to take place in November 2014.

**5. CONCLUSIONS**

The work of the HSTI during the past three years from 2010 to 2013 was characterized by the introduction of HSTI as a new initiative within the United Nations Programme on Space Applications, and making the Initiative known to the Member States and various sectors. The Initiative has aimed to provide a platform for nations to make use of it on the basis of their interest in participating in increased efforts in human space exploration, and/or in enhancing capacity on human space technology and its applications.

During its initial phase, HSTI defined the three pillars of activities - international cooperation, outreach and capacity-building. HSTI recognizes that human space exploration could be regarded as a common goal of humanity and that all nations should be encouraged to become involved in understanding and defining the common goals and important benefits of such global efforts. HSTI started working with the ISS partner agencies to extend the research and educational benefits of the ISS to more people on Earth. HSTI has also engaged in dialogue with China to collaborate with its space station programme. HSTI will continue to strengthen its pillar of international cooperation for more nations to engage in advancing human space exploration and its related activities.

HSTI has conducted mainly two activities for the outreach pillar. They are the United Nations/Malaysia Expert Meeting on Human Space Technology in 2011 and the United Nations/China Workshop on Human Space Technology in 2013. Experts from 22 countries participate in the 2011 Expert Meeting; experts from 31 countries participate in the 2013 Workshop. A total of 38 countries participated in the activities conducted under HSTI. This proves that human space exploration and its related activities have become truly global undertakings. However there are still many nations which are barely aware of the benefits of human space technology and miss opportunities in participating in exploration and uses of outer space. HSTI will continue to be an advocate to raise the awareness of Member States and the general public about the importance of human space exploration and its societal benefits.

In order for nations to join together in efforts for future space exploration, it is essential that the nations have knowledge, expertise and resources to participate effectively. Hence, building and enhancing technological and scientific capacity of space experts is crucial for non-space-faring countries. HSTI initiated its capacity-building activities by introducing the Zero-Gravity Instrument Project (ZGIP) as an educational and research project and the Drop Tower Experiment Series (DropTES) as a fellowship programme. Both activities focus on introducing microgravity science education and research, particularly in developing countries.

8th IAA Symposium on the future of space exploration: towards the starsFrom 2014 to 2016, HSTI will further strengthen its three pillars and strive to bring the benefits of human space activities to all and to bring more nations together for that endeavour and thus create new opportunities for international cooperation.

**ACKNOWLEDGMENTS AND DISCLAIMER**

We are grateful to all who have joined the activities of HSTI, especially to those who took part in the meetings and seminars from 2011 to 2013 and those who provided constructive comments in the process of the HSTI implementation.

The views expressed herein are those of the authors and do not necessarily reflect the views of the United Nations.